\documentclass[]{aastex631}

\usepackage{appendix}

\usepackage{enumitem}
\usepackage{comment}
\usepackage{booktabs}
\usepackage{threeparttable}
\usepackage{xcolor}
\usepackage{color, colortbl}
\usepackage{array,multirow,graphicx}
\usepackage [english]{babel}
\usepackage [autostyle, english = american]{csquotes}
\MakeOuterQuote{"}

\usepackage{amsmath,amssymb}

\DeclareMathOperator{\EX}{\mathbb{E}}

\def\s{\phantom}

\def\amin{\ifmmode^{\prime}\else$^{\prime}$\fi}
\def\asec{\ifmmode^{\prime\prime}\else$^{\prime\prime}$\fi}

\def\simgt{\lower.5ex\hbox{$\; \buildrel > \over \sim \;$}}
\def\simlt{\lower.5ex\hbox{$\; \buildrel < \over \sim \;$}}

\newcommand\chandra{{\it Chandra}}
\newcommand\xmm{{\it XMM-Newton}}

\newcommand\integral{{\it INTEGRAL}}

\newcommand\nustar{{\it NuSTAR}}

\newcommand{\fluxcgs}{ergs~s$^{-1}$~cm$^{-2}$}
\newcommand{\lumcgs}{ergs~s$^{-1}$}

\shorttitle{Galactic Center HMXBs}
\shortauthors{Mandel et al.}

\begin{document}

\title{
Hunting for High-Mass X-ray Binaries in the Galactic Center with NuSTAR
}

\author[0000-0002-6126-7409]{Shifra Mandel}
\affiliation{Columbia Astrophysics Laboratory, Columbia University, New York, NY 10027, USA}

\author[0000-0001-6470-6553]{Julian Gerber}
\affiliation{Columbia Astrophysics Laboratory, Columbia University, New York, NY 10027, USA}

\author[0000-0002-9709-5389]{Kaya Mori}
\affiliation{Columbia Astrophysics Laboratory, Columbia University, New York, NY 10027, USA}

\author[0009-0009-8275-6111]{Ceaser Stringfield}
\affiliation{Columbia Astrophysics Laboratory, Columbia University, New York, NY 10027, USA}

\author[0009-0009-3206-7326]{Mabel  Peñaherrera}
\affiliation{Columbia Astrophysics Laboratory, Columbia University, New York, NY 10027, USA}

\author[0000-0002-3681-145X]{Charles J. Hailey}
\affiliation{Columbia Astrophysics Laboratory, Columbia University, New York, NY 10027, USA}

\author[0009-0006-7033-6943]{Alan Du}
\affiliation{Division of Physics, Mathematics and Astronomy, California Institute of Technology, Pasadena, CA 91125, USA}

\author[0000-0002-1323-5314]{Jonathan Grindlay}
\affiliation{Harvard-Smithsonian Center for Astrophysics, Cambridge, MA 02138, USA}

\author[0000-0002-6089-5390]{JaeSub Hong}
\affiliation{Harvard-Smithsonian Center for Astrophysics, Cambridge, MA 02138, USA}

\author[0000-0003-0293-3608]{Gabriele Ponti} 
\affiliation{INAF - Osservatorio Astronomico di Brera, Merate, Italy}
\affiliation{Max-Planck-Institut für extraterrestrische Physik, Garching, Germany}

\author[0000-0001-5506-9855]{John A. Tomsick}
\affiliation{Space Sciences Laboratory, 7 Gauss Way, University of California, Berkeley, CA 94720-7450, USA}

\author[0000-0003-0746-795X]{Maureen van den Berg}
\affiliation{Harvard-Smithsonian Center for Astrophysics, Cambridge, MA 02138, USA}

\correspondingauthor{Shifra Mandel}
\email{ss5018@columbia.edu}

\correspondingauthor{Julian Gerber}
\email{jmg2369@columbia.edu}


\begin{abstract}
The central $2\times0.8$ deg$^2$ region of our Galaxy contains $\sim10,000$ X-ray point sources that were detected by a series of \chandra\ observations 
over the last two decades. However, the limited bandpass of \chandra\ below 8 keV hampered their spectroscopic classification. In 2016, the initial \nustar\ Galactic center (GC) survey detected 77 X-ray sources above 10 keV \citep{Hong2016}. The hard X-ray detections indicate magnetic cataclysmic variables (CVs), low-mass X-ray binaries (LMXBs), high-mass X-ray binaries (HMXBs), or even pulsars. The possibility of HMXB detections is particularly interesting 
given the dearth of identified HMXBs in the GC.
We conducted a search for bright ($K_s\simlt16$ mag) near-infrared (NIR) counterparts to the hard X-ray sources -- utilizing their \chandra\ positions -- in order to identify HMXB candidates.  
We identified seven \nustar\ sources with NIR counterpart candidates whose 
magnitudes are consistent with HMXBs at the GC.  
We assessed the likelihood of random association for these seven sources and determined that two have a non-random association with a probability exceeding $99.98\%$, making them strong HMXB candidates.  
We analyzed broadband \nustar, \chandra\ and \xmm\ spectral data for these two  
candidates, one of which was previously identified as a red supergiant.  We find that the X-ray spectra are consistent with HMXBs.  If confirmed through follow-up NIR spectroscopic studies, our findings will open a new window into our understanding of the intrinsic luminosity distribution of HMXBs in our Galaxy in general and the GC HMXB population in particular.  

\end{abstract}


\section{Introduction}\label{sec:intro}

High-mass X-Ray binaries (HMXBs) are a subclass of X-ray binaries containing a compact object (either a black hole or neutron star) and companion star of mass $\simgt5 M_{\odot}$ \citep{Fortin2023}. In some HMXBs, the companion -- or "donor" star -- is an OB type supergiant; accretion onto the compact object typically occurs through the dense stellar wind ejected from the massive donor, or, more rarely, through Roche lobe overflow \citep{Chaty2011}.  
A substantial fraction of the known Galactic HMXBs contain a Be-type donor, which is surrounded by a dense, circumstellar decretion disk \citep{Chaty2011, Fortin2023}.  
In these systems, mass transfer occurs when the compact object interacts with the circumstellar disk.  Apart from their significance for our understanding of the evolution of massive stars, HMXBs are important for their role as progenitors of double compact object systems, which result in 
gravitational wave sources \citep{Belczynski2016, Gies2019}.  HMXBs are also relevant in the study of the Epoch of Reionization (EoR), the period of the early universe when the first stars and galaxies formed; X-ray emission from HMXBs are believed to be the dominant contributors to the EoR at high redshifts ($z\simgt6$--8; \citet{Fragos2013}).

Since its launch in 2002, the International Gamma Ray Astrophysics Laboratory (\integral) has been successful in expanding the number of known HMXBs. This is due to the fact that HMXBs are copious emitters of hard X-rays. In the majority of cases, where an HMXB contains an accreting pulsar, X-rays are produced when inflowing plasma is forced to travel along the neutron star's magnetic field lines, reaching close to the speed of light as it falls onto the accretion columns above the pulsar's magnetic poles. The X-ray spectrum is usually characterized by a hard power-law -- resulting from Comptonization -- with photon index that can range between $\Gamma=0.3-2$, as well as a high energy ($7-30$ keV) exponential cutoff \citep{Filippova2005, Walter2015}.  
Other signatures that sometimes appear in accreting pulsars include high-energy cyclotron resonance scattering features (CRSF) \citep{Coburn2002, Caballero2012}.
The most recent all-sky \integral\ survey utilized 17 years of \integral\ data to identify a total of 115 Galactic HMXBs \citep{Krivonos2022}. 
Additional HMXBs and HMXB candidates were identified in \citet{Fortin2023}, bringing the total to 152 -- although this distribution appears strongly biased towards brighter ($L_{\rm X} \simgt10^{34}$ \lumcgs) systems, raising the question of whether a larger population of fainter HMXBs has gone undetected.  

Within the Galactic center (GC), the Central Molecular Zone (CMZ),
with its overabundance of compact objects in the central parsec \citep{Hailey2018, Mori2021}, strong recent star formation activity \citep{Nogueras-Lara2020}, 
and top-heavy IMF \citep{Lu2013, Hosek2019}, is the perfect environment to host an HMXB.  
Of all the known Galactic HMXBs, however, only one has hitherto been found within $\sim100$ pc of Sgr A* \citep{Liu2006, DeWitt2010, Bird2016, Gottlieb2020, Fortin2023}, a region that hosts $\sim10-15\%$ of all Galactic LMXBs \citep{Mori2021, Fortin2024}.
Furthermore, the X-ray luminosity function of Galactic HMXBs is only well studied down to $\sim10^{34}$ \lumcgs \citep{Lutovinov2013}. A 2017 
\nustar\ Serendipitous Survey demonstrated how the improved sensitivity of \nustar\ can be used to detect intrinsically faint HMXBs at greater distances \citep{Tomsick2017}. 
For example, 
the NuSTAR source J105008-5958.8 was identified as an HMXB candidate with an 
X-ray luminosity of $1.8-5.9\times10^{32}$ \lumcgs \citep{Tomsick2017}, 
about two orders of magnitude fainter than the lower end cut-off of the previously 
known HMXB luminosity function. The existence of a faint HMXB candidate suggests that the absence of faint HMXBs from the known population may be 
a result of sensitivity limits on the instrumentation rather than a reflection of the true population of HMXBs.  
Understanding the lower end of the HMXB luminosity function is important to fully understand emission processes at lower accretion rates \citep{Fornasini2016PhDT}. 
Telescopes such as \chandra, \nustar, and \xmm\ can have sensitivities up to four orders of magnitude fainter than that of \integral. These telescopes are therefore more likely to detect this possible faint population of HMXBs, particularly at distances as far as the GC.

\begin{figure}
    \centering
    \includegraphics[width=15cm]{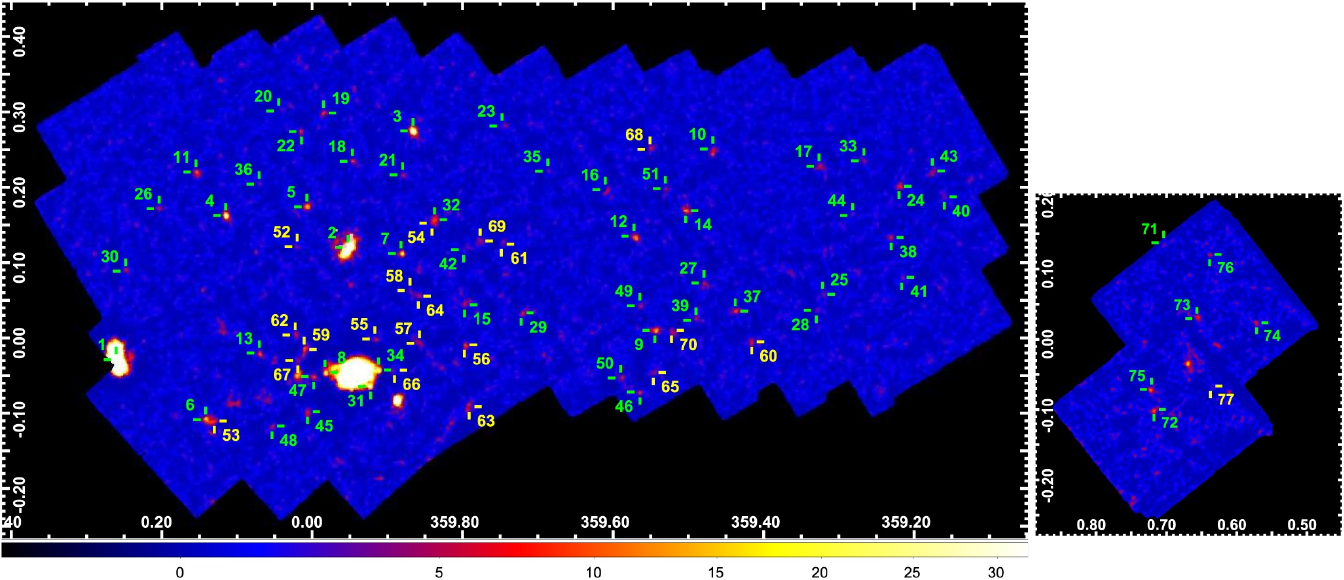}
    \caption{10--40 keV \nustar\ trial map of the GC showing the 77 point sources, most of which were detected in the hard ($>10$ keV) X-ray band.  Colors are 
    scaled by trial number, a measure of source significance.   From figure 3 in \citet{Hong2016}.\label{fig:ngp}}
\end{figure}

Previous studies have utilized follow-up \nustar, \chandra, and \xmm\ observations in order to better understand 
individual 
HMXBs. For example, IGR J18214-1318 was first discovered in the \citet{Bird2006} \integral\ catalog. Follow-up \chandra\ observations conducted by \citet{Tomsick2008} allowed for a greater positional accuracy and subsequent identification of a high-mass optical counterpart \citep{Butler2009}. Additional \nustar\ and \xmm\ observations of the source allowed for greater insight into the compact object within the HMXB \citep{Fornasini2017}. In this paper, we continue this strategy of using \nustar, \chandra, and \xmm\ observations to improve our understanding of HMXB candidates. Furthermore, the recent release of improved near-infrared (NIR) surveys such as 
GALACTICNUCLEUS and VVV DR5 \citep{GALACTICNUCLEUS2019, vvv2023}
make this an optimal time to search for NIR counterparts within the crowded CMZ region. Earlier X-ray--Optical/NIR surveys using only \chandra\ and/or \xmm\ X-ray data, such as \citet{vandenBerg2009, Mauerhan2009,  Greiss2014, Nebot2015} and \citet{Morihana2022}, did not result in the discovery of new HMXBs. 

We evaluated the 77 sources detected in the \nustar\ Hard X-Ray Survey of the GC (\citet{Hong2016}; see Figure \ref{fig:ngp}) to search for faint HMXB candidates. We queried various NIR catalogs for bright NIR counterparts to each source and estimated the expected number of random matches for potential counterparts (Section \ref{sec:ccorr}).  
For the strongest HMXB candidates, we 
conducted joint X-ray spectral analysis using archival data from \chandra, \nustar, and \xmm\ (Section \ref{sec:spectra}). 
Finally, in Section \ref{sec:discuss} we discuss the potential natures of these sources based on the spectral results as well as further implications for the study of faint HMXBs in the CMZ region of the GC. \\

\section{Cross-correlation of X-Ray and Near-infrared Sources}\label{sec:ccorr}

HMXBs are distinguished by their high-mass donors, stars of O or B type that are highly luminous in the 
NIR and optical bands ($>500\,L_{\odot}$).  As the GC is virtually opaque in the optical due to high extinction ($A_V\sim30$ mag), we turned to NIR data to search for potential NIR counterparts to our X-ray sources.  
Assuming a maximum absolute magnitude for an O/B type HMXB donor star of $M_K \sim-0.3$ \citep{Pecaut2013}, 
a distance of $\sim8$ kpc to the GC \citep{Leung2023}, and mean absorption $A_K\simgt1.5$ mag \citep{2021Nogueras}, we set a threshold for $K_s$-band apparent magnitude of $\leq16$ for NIR HMXB counterpart candidates.  We also required $H-K_s > 1.3$ mag, consistent with an intrinsic color of $-0.1\leq H-K_s \leq0$ and the extinction expected at the GC distance \citep{2021Nogueras}.

To identify NIR counterpart candidates for our X-ray sources, we searched the most sensitive NIR survey catalogs of the GC for bright ($K_s < 16$ mag), reddened ($H-K_s > 1.3$ mag) sources that overlap the positions of the unidentified sources -- identified by their NuSTAR GC Point (NGP) source ID -- detected in the \nustar\ hard X-Ray survey of the GC \citep{Hong2016}.  
We define a NIR source as overlapping an X-ray source if the position centroid offset between the two sources is smaller than the sum of their position uncertainties.  
Since \nustar's angular resolution ($18\asec$ FWHM) is insufficient to constrain the source positions, we started by cross-matching each of the \nustar\ sources with the most recent \chandra\ source catalog, CSC2.1 \citep{Evans2024}.  Where multiple \chandra\ sources were found within $10\asec$ of the \nustar\ position centroid (as given in table 2 of \citet{Hong2016}), we attempted to determine the best match by comparing the \nustar\ $3-10$ keV flux and spectral index to those of the overlapping \chandra\ sources.  Once a \chandra\ counterpart was identified, we adopted the \chandra\ positions and the 95\% position uncertainties published in CSC2.1 \citep{Evans2024} for all subsequent NIR counterpart searches.

\subsection{NIR Catalogs: VVV and GALACTICNUCLEUS}

\begin{figure}
    \centering
    \includegraphics[width=\textwidth]{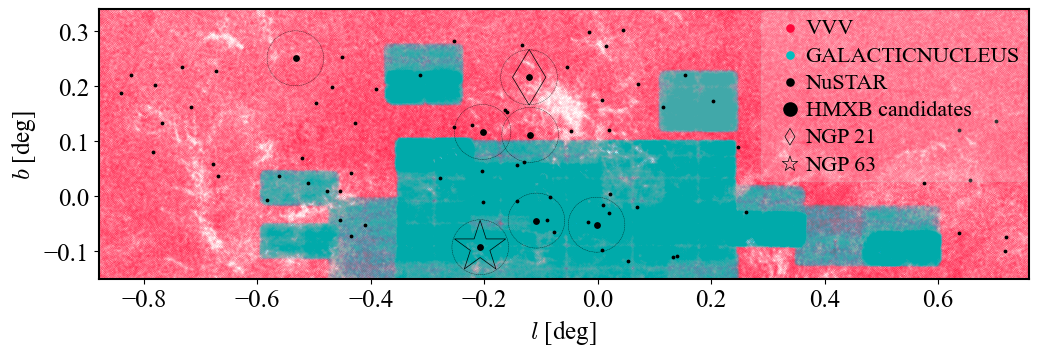}
    \caption{\nustar\ hard X-ray sources (black) were cross-matched to the VVV (red) and GALACTICNUCLEUS (cyan) NIR catalogs to identify possible counterparts.  HMXB candidates have larger markers, and are surrounded by dashed circles of radius 2\arcmin.  The two strongest candidates, NGP 21 and NGP 63, are also marked with a diamond and star, respectively.}\label{fig:reg}
\end{figure}

The VISTA Variables in the Via Lactea Survey (hereafter, VVV) is the deepest
NIR survey that uniformly covers the central degrees of the Galaxy \citep{vvv2023}.  VVV is saturated at magnitudes brighter than $K_s\sim11$ \citep{FerreiraLopes2020}, and completeness starts to fall off near $K_s\simgt14$ mag (Figure \ref{fig:den}) due to crowding \citep{GALACTICNUCLEUS2019}.  We utilized the most recent data release, VVV DR5, in our initial search for NIR counterparts to the hard X-ray sources detected by \nustar.  We find overlapping bright VVV sources for five \nustar\ sources.  Each of those VVV sources have $K_s$ magnitudes and $H-K_s$ colors consistent with B-type stars at the GC.  However, we note that at the GC, NIR colors are strongly dominated by extinction, hence we cannot determine based on $H-K_s$ color alone whether a source is an O/B-type star.

Within the nuclear stellar disk (NSD), an even deeper NIR survey yielded the GALACTICNUCLEUS source catalog, which is 80\% complete at $K_s\sim16.3$ mag \citep{2021Nogueras}.  The latter overlaps part of the \nustar\ GC survey (Figure \ref{fig:reg}).  Searching this catalog, we found bright NIR sources that overlap the position uncertainty regions 
for three of the \nustar\ sources, including NGP 63, for which we had already identified a counterpart in the VVV data.  The similarity in brightness and position of the GALACTICNUCLEUS counterpart candidate relative to the VVV counterpart candidate for NGP 63 indicates that they are in fact the same source.  The two other candidate counterparts found in the GALACTICNUCLEUS data, which are considerably fainter, were not detected in the VVV survey, likely due to the latter's poorer depth; the GALACTICNUCLEUS survey's better angular resolution renders it less susceptible to the crowding that plagues its VVV counterpart at magnitudes $\simgt14$.  
The $K_s-$band magnitudes and $H-K_s$ colors of all three GALACTICNUCLEUS counterpart candidates are consistent with a B-type star at the GC (but as mentioned above, NIR $H-K_s$ color is by itself insufficient to distinguish O/B-type stars from other GC sources).

\begin{figure}
\centering
\includegraphics[width=\textwidth]{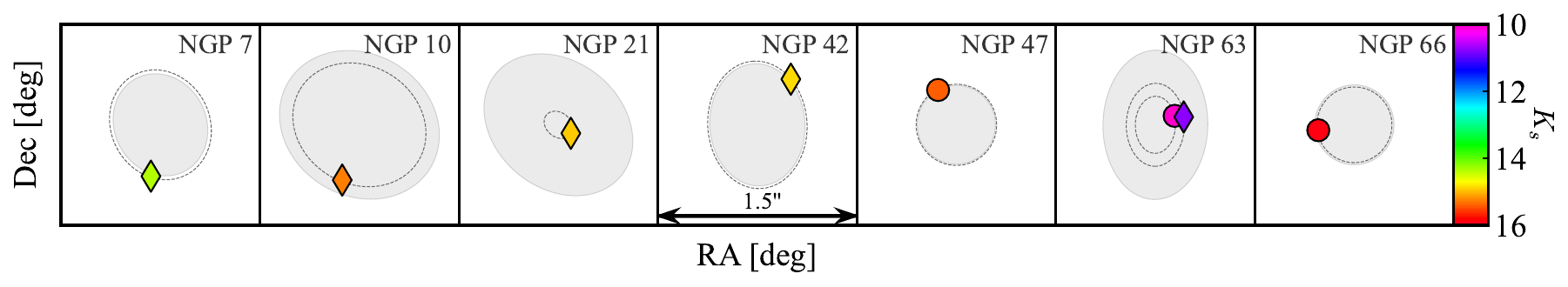}
\caption{Relative positions of potential NIR counterparts for HMXB candidates.  Gray ellipses denote the \chandra\ 95\% uncertainty regions.  Colored markers correspond to GALACTICNUCLEUS (circles) or VVV (diamonds) NIR counterpart candidates.  Dark dotted ellipses mark the area $A_{\phi}$ used to calculate the expected number of random matches listed in Table \ref{tab:hmxbcands1} (cf. Section \ref{subsec: rprob1}).  We believe that the GALACTICNUCLEUS and VVV sources overlapping with NGP 63 are one and the same, given their small spatial offset and similar magnitudes.}
\label{fig:csc}
\end{figure}

In total, seven \nustar\ sources were found to be plausible HMXB candidates given their potential associations with bright, reddened NIR counterparts.  Figure \ref{fig:csc} shows the 95\% uncertainty region (grey shaded ellipse) associated with the \chandra\ coordinates for each HMXB candidate, overlaid with the NIR counterpart candidates (colored markers).  Table \ref{tab:hmxbcands1} lists the seven (VVV+GALACTICNUCLEUS) NIR HMXB candidate counterparts, along with their $K_s$ magnitudes and colors.  Hereafter we refer to the $K_s$ photometric band simply as $K$ to avoid confusion as other subscripts are introduced.

We find that although the GALACTICNUCLEUS survey is significantly deeper
overall compared to VVV, its depth is somewhat inconsistent across different fields, a result of variations in observing conditions between the single-epoch pointings \citep{GALACTICNUCLEUS2019}.  By contrast, VVV appears more uniform and has the additional advantage of overlapping the entire \nustar\ survey region, making it more suitable for the expected number of random matches studies in Sections \ref{subsec: rprob1} and \ref{subsec:mmprob}.

\subsection{Random Match Probability Estimates}\label{subsec: rprob1}

\begin{figure}
\centering
\includegraphics[width=0.48\textwidth]{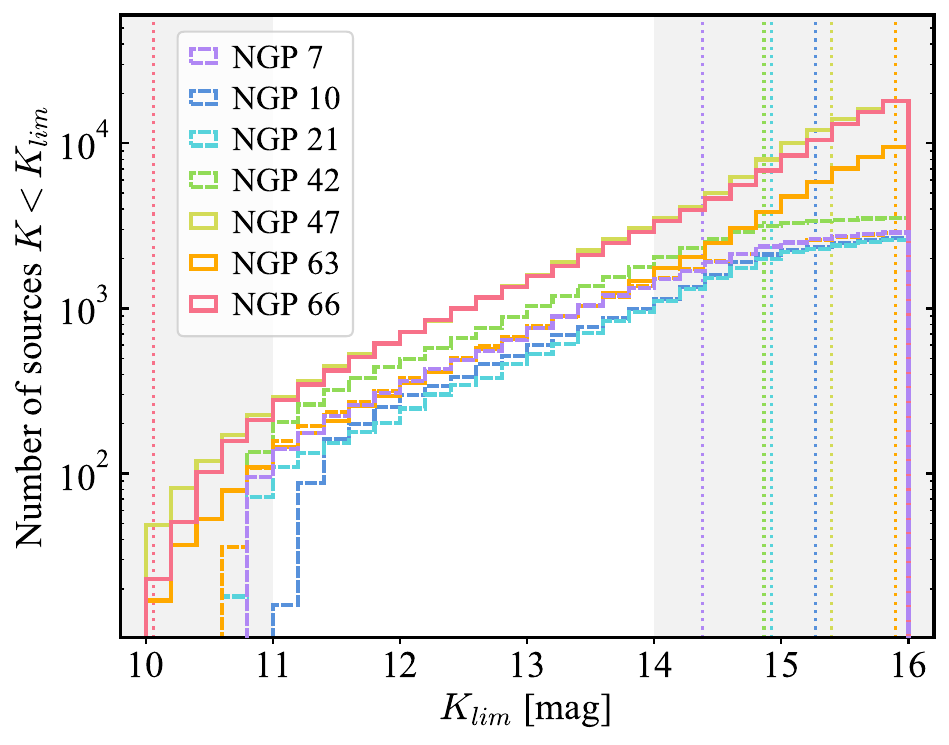}
\includegraphics[width=0.49\textwidth]{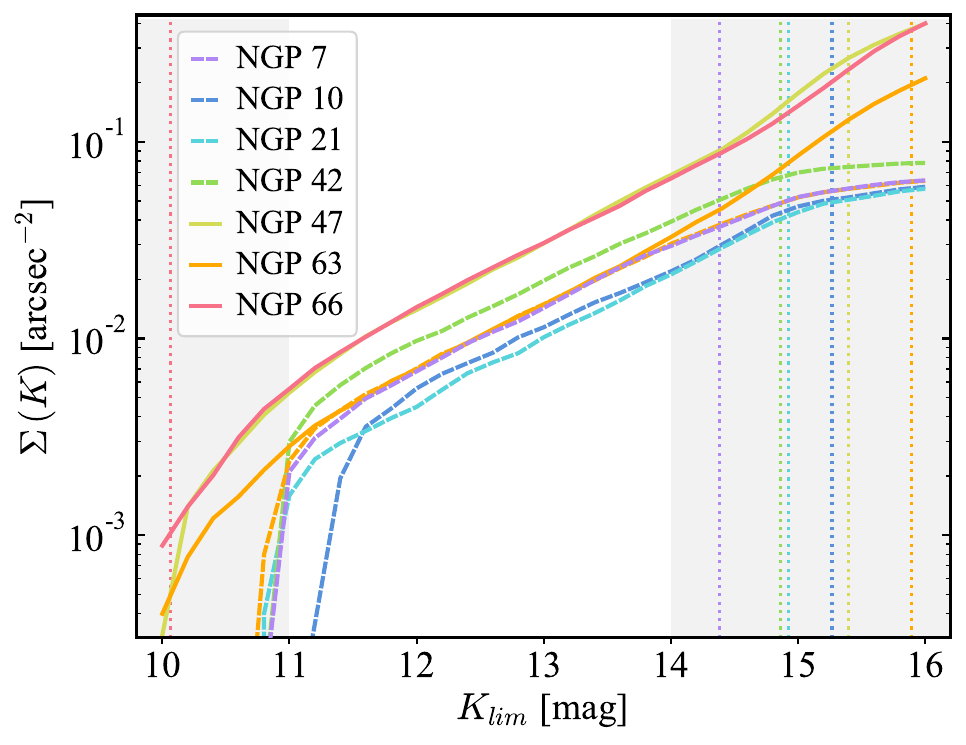}
\caption{Left: Cumulative histograms showing the magnitude distributions within a 2\arcmin\ region around each HMXB candidate.  Right: surface density profiles $\Sigma(K)$ within the same 2\arcmin\ region around each HMXB candidate.  Solid lines correspond to GALACTICNUCLEUS data, dash-dotted lines to VVV data.  Vertical dotted lines denote the magnitude of the corresponding NIR counterpart candidate to each X-ray source.  The shaded areas mark the magnitude ranges where the VVV survey is incomplete.  
See Section \ref{sec:AC} in the appendix for more.}
\label{fig:den}
\end{figure}

In order to determine the probability that the NIR HMXB candidate counterparts we identified are truly associated with the X-ray sources whose positions they overlap, we must first evaluate the 
number of NIR sources that are expected to \emph{randomly} overlap with the X-ray source position.  This expected value is 
influenced by several factors, primarily the local surface density of NIR sources.  
As Figure \ref{fig:den} (left) illustrates, faint sources are much more abundant than brighter ones, hence source density is inversely related to luminosity and increases exponentially with magnitude -- up to the limit of a catalog's completeness, where the number of sources detected falls off due to source confusion, a result of crowding (but NOT intrinsic population decline at fainter magnitudes, given the depth of these NIR surveys).  
This phenomenon is highlighted in Figure \ref{fig:den} (right), which shows the surface density $\Sigma(K)$ of NIR sources below a range of magnitude thresholds $K_{lim}$, within the local ($r<2\arcmin$) region surrounding each HMXB candidate.

Another important factor in determining the probability of a random match occurring between an unrelated X-ray and NIR source is the distance between them; we define the offset between the centroids of the X-ray and NIR source positions as $\phi$, which is inversely correlated with the expected number of random matches.  The offset $\phi$ is especially significant in the context of the uncertainty in the X-ray and NIR source positions:  $\sigma^2=\sigma_{\rm X}^2+\sigma_{\rm IR}^2$, where $\sigma_{\rm X}$ is the uncertainty associated with the X-ray position and $\sigma_{\rm IR}$ is the uncertainty in the NIR source position.  We can then describe the probability 
distribution for the offset $\phi$ between an X-ray source's position centroid and that of a NIR counterpart (with a "real" association) as 
\begin{equation} \label{eq:phi}
    p(\phi)=\frac{1}{2 \pi \sigma^2}exp(\frac{-\phi^2}{2 \sigma^2})
\end{equation}
assuming a Gaussian form for the position uncertainty \citep{Evans2024}.  Figure \ref{fig:pdf} in the appendix illustrates the probability distributions $p(\phi)$ for a range of $\sigma$ values.  The probability density $p(\phi)$ is highest within $\phi<\sigma$ (dotted line) and falls off very rapidly beyond the 95\% ($\sim2\sigma$) limit (dashed line, corresponding to \chandra\ uncertainty level; we approximate $\sigma \sim \sigma_{\rm X}$, since $\sigma_{\rm X}>>\sigma_{\rm IR}$).  Section \ref{subsec:mmprob} briefly covers the probability of a "true" counterpart being located outside of the 95\% uncertainty region, but overall, we focus our efforts on the matches located at $\phi\simlt2\sigma$.  Since the \chandra\ 95\% uncertainty regions are defined by an ellipse, we denote an offset "region" $A_{\phi}$ by an ellipse with the same inclination and eccentricity $e$ as the X-ray position uncertainty, 
scaled such that the NIR source is at its edge, as illustrated by the dark outlines in Figure \ref{fig:csc}.  Appendix \ref{sec:AB} describes the process of defining $A_{\phi}$ in more detail.

For NIR sources that fall beyond the completeness limit of their respective catalogs, there is an additional step we need to take before we can determine the expected number of random matches $\EX_{\rm rand}$.  That is, we need to estimate the level of completeness $\epsilon_i$ at the magnitude $K_i$ corresponding to the NIR source.  Since the GALACTICNUCLEUS catalog is 80\% complete at $K=16.3$ mag \citep{2021Nogueras}, this issue primarily affects the VVV sources.  As Figure \ref{fig:den} shows, the number of sources in the VVV survey begins to fall off at magnitudes fainter than $K \sim14$ due to crowding  (cf. also figure 12 in \citet{GALACTICNUCLEUS2019}, which compares the VVV and GALACTICNUCLEUS luminosity functions).  Since $\epsilon_i$ varied slightly over different regions, we estimated the projected number of sources around each HMXB candidate individually.  As Figure \ref{fig:comp} in the appendix illustrates, we generated the source population estimates by fitting a line to the cumulative logarithmic distribution \emph{within the magnitude range} where the survey is complete, i.e. $\epsilon\simeq100\%$.  We then extrapolated that line to derive the estimated source populations at fainter magnitudes\footnote{While this derivation does not explicitly account for non-exponential features in the magnitude distribution (like the Red Clump bump), it is nevertheless a good approximation of the local magnitude distribution based on the empirical data.}.  We define the fractional completeness $\epsilon_i$ at magnitude $K_i$ as $N_{\rm obs}(K_i)/N_{\rm proj}(K_i)$, where $N_{\rm obs}(K_i)$ is the number of sources observed at $K < K_i$, and $N_{\rm proj}(K_i)$ is the number of sources projected at $K < K_i$.  More details on the completeness estimates can be found in Appendix \ref{sec:AC}.

We can then define the expected number of NIR sources of magnitude $K_i$ randomly overlapping 
an X-ray source as 
\begin{equation} \label{eq:rmp}
    \EX_{\rm rand}(K_i) = \Sigma(K_i)\times A_{\phi}/\epsilon_i
\end{equation}
where $A_{\phi}$ is the size of the aforementioned elliptical region, centered on the X-ray position, that encompasses the NIR counterpart candidate, defined by the semi-major axis $a$ and semi-minor axis $b$:   $A_{\phi}=\pi ab$.  

\setlength{\tabcolsep}{0pt}
\begin{table}
\begin{center}
\caption{HMXB candidates.  We list the \nustar\ ID, \chandra\ counterpart, NIR survey ("GN" denotes GALACTICNUCLEUS), NIR counterpart candidate coordinates, NIR counterpart candidate magnitude $K_i$, NIR color $H-K$, the local surface density $\Sigma(K_i)$ of NIR sources at magnitude brighter than $K_i$, survey completeness $\epsilon_i$ at $K_i$ (for VVV), and finally, the expected number of random matches $\EX_{\rm rand}(K_i)$, as defined in Section \ref{subsec: rprob1}.  
The two strongest HMXB candidates are highlighted in grey.  }
\vspace{-0.3cm}
\begin{tabular}{ l c c c c l l c c c } 
 \hline\hline
 Source $\;\;$ & $\;\;$ \chandra\ ID $\;\;$ &  NIR survey  & $\;\;$ $l$ $\,$  & $\,$ $b$  $\;\;$ & $\;\;$ $K_i\,^a$ $\;\;$ & $\;\;$ $H-K$  $\;\;$ & $\;\;$ $\Sigma(K_i)$ $\;\;$ & $\;\;$ $\epsilon_i$ $\;\;$ & $\;\;$ $\EX_{\rm rand}(K_i)$  \\ [-2pt]
 & $\;\;$ 2CXO $\;\;$ & $\;\;$ & $\;\;$ [deg] $\,$ & $\,$ [deg] $\;\;$ & $\;\;$ [mag] $\;\;$ & $\;\;$ [mag] $\;\;$ & $\;\;$ [arcsec$^{-2}$] $\;\;$ & $\;\;$ [\%] $\;\;$ & $\;\;$ [\%] \\[0.5ex]
 \hline
NGP 7 & $\;\;$ J174454.1-285841 $\;\;$ & VVV & $\;\;$ 359.88207 
$\,$ & $\,$ 0.11191 
$\;\;$ & $\;\;$ 14.4 
$\;\;$ &  $\;\;$ 1.92 $\;\;$ & $\;\;$ $3.68\times10^{-2}$ $\;\;$  & $\;\;$ 81.2 $\;\;$  & $\;\;$ 2.17  \\ 
NGP 10 & $\;\;$ J174321.9-291530 $\;\;$ & VVV & $\;\;$ 359.46745 $\,$ & $\,$ 0.25105 
$\;\;$ & $\;\;$ 15.3 $\;\;$ &  $\;\;$ 1.55 $\;\;$ & $\;\;$ $5.07\times10^{-2}$ $\;\;$ & $\;\;$ 58.7 $\;\;$ & $\;\;$ 6.05  \\ 
\rowcolor[HTML]{E5E7E9}
NGP 21 & $\;\;$ J174429.4-285531 $\;\;$ & VVV &  $\;\;$ 
359.88003 $\,$ & $\,$ 0.21638 $\;\;$ & $\;\;$ 14.9 $\;\;$ & $\;\;$ 1.58 $\;\;$ & $\;\;$ $4.19\times10^{-2}$ $\;\;$ & $\;\;$ 96.7 $\;\;$ & $\;\;$ 0.14  \\ 
NGP 42 & $\;\;$ J174441.0-290248 $\;\;$ & VVV & $\;\;$ 
359.79877 $\,$ & $\,$ 0.11690 $\;\;$ & $\;\;$ 14.9  &  $\;\;$ 2.08$^b$ & $\;\;$ $6.58\times10^{-2}$ $\;\;$ & $\;\;$ 82.2 $\;\;$ & $\;\;$ 4.36  \\ 
NGP 47 & $\;\;$ J174548.9-285751 $\;\;$ & GN & $\;\;$ 359.99830 $\,$ & $ $ -0.05123 $\;\;$ & $\;\;$ 15.4 $\;\;$ & $\;\;$ 2.39 $\;\;$ & $\;\;$ $2.67\times10^{-1}$ $\;\;$ & $\;\;$ -- $\;\;$ & $\;\;$ 7.51  \\ 
\rowcolor[HTML]{E5E7E9}
NGP 63 & $\;\;$ J174528.7-290942 $\;\;$ & VVV & $\;\;$ 359.79140 $\,$ & $ $ -0.09149 $\;\;$ & $\;\;$ 10.8$^c$ & $\;\;$ 1.81$^c$ & $\;\;$ $5.97\times10^{-4}$ $\;\;$ & $\;\;$ N/A $\;\;$ & $\;\;$ $1.24\times10^{-2}$  \\ 
\rowcolor[HTML]{E5E7E9}
 & $\;\;$  $\;\;$ & GN & $\;\;$ 359.79139 $\,$ & $ $ -0.09147 $\;\;$ & $\;\;$ 10.1  & $\;\;$ 2.76  & $\;\;$ $5.08\times10^{-4}$ $\;\;$ & $\;\;$ -- $\;\;$ & $\;\;$ $5.01\times10^{-3}$  \\ 
NGP 66 & $\;\;$ J174531.5-290307 $\;\;$ & GN & $\;\;$ 359.89041 $\,$ & $ $ -0.04278 $\;\;$ & $\;\;$ 15.9  & $\;\;$ 2.12  & $\;\;$ $3.74\times10^{-1}$ $\;\;$ & $\;\;$ -- $\;\;$ & $\;\;$ 9.12  \\ 
 \hline
\end{tabular}   
\end{center}
\vspace{-0.2cm}
$^a$Typical uncertainties in $K$ are approximately 0.1 mag for VVV and 0.01 mag in the GALACTICNUCLEUS catalog. \\
$^b$The VVV data did not include $H-$band magnitude for this source.  We adopted the color data from the UKIDSS-DR6 catalog \citep{UKIDSSDR6}. \\
$^c$
The discrepancy between the VVV and GALACTICNUCLEUS magnitudes/colors (and the corresponding $\Sigma$ and $\EX_{\rm rand}$) 
could be due to either intrinsic variability \citep{Gottlieb2020} or saturation in the VVV survey at $K<11$ mag (\citet{FerreiraLopes2020}; cf. also Figure \ref{fig:comp}). \\
\label{tab:hmxbcands1}
\end{table}

Table \ref{tab:hmxbcands1} lists the expected number of random matches $\EX_{\rm rand}$ for each of the seven  HMXB candidates, along with the surface density $\Sigma(K_i)$ and completeness $\epsilon_i$ at magnitude $K_i$ corresponding to that of the NIR counterpart candidate.  Two of the sources, NGP 21 and NGP 63 (highlighted), have random associations ruled out beyond the $3\sigma$ level. \\

\section{X-Ray Spectral Analysis}\label{sec:spectra}

\subsection{Data Reduction}
We conducted joint X-ray spectral analysis on NGP 21 and NGP 63, the two \nustar\ sources with the highest probability ($>3\sigma$) of a "real" bright NIR counterpart. We utilized archival data from \nustar, \chandra, and \xmm\ where the source was clearly visible while discarding observations and spectra where the background dominated.

\subsubsection{\chandra}
To obtain \chandra\ spectra, we queried CSC2.1 for source spectra, background spectra, and response files for the \chandra\ counterparts of NGP 21 and NGP 63. After downloading all the available data, we discarded any spectra dominated by background. Specifically, we removed spectra where the source counts accounted for less than 50\% of the total counts. We then merged the remaining spectra and response files using CIAO's {\tt combine\_spectra} tool. Finally, we binned the merged spectrum by a minimum signal-to-noise (SNR) ratio of two using HEASoft's {\tt ftgrouppha} command. We fit the \chandra\ spectra in the 2 -- 8 keV range. This energy range was chosen since, below 2 keV, \chandra\ suffers some attenuation of photons due to a buildup of contamination on the ACIS detectors. \citep{ODell2017} Likewise, above 8 keV, \chandra's effective area decreases significantly. \citep{Zhao2004}

\subsubsection{\xmm}
NGP 21 only had one \xmm\ observation (ObsID 0506291201) available within $8^\prime$ of its \chandra\ coordinates. However, NGP 21 was not visible in this observation's 
image. We therefore did not use \xmm\ data in the spectral analysis of NGP 21.

NGP 63, on the other hand, had two XMM observations (ObsID 0694641001 and 0862470401) within $8^\prime$. In both observations, the source region was directly between two detector chips in the PN images. Therefore, only the MOS 1 and MOS 2 cameras were used in each observation

To reduce the EPIC/MOS data, we used the XMM-Newton Science Analysis Software (SAS) version 21.0.0. We first ran the SAS {\tt emchain} script to generate the processed event lists for MOS 1 and 2. We filtered the MOS 1/2 data with the expression (PATTERN $<$= 12)\&\&(PI in [200:12000])\&\&\#XMMEA\_EM. In order to account for soft proton flares, we constructed light curves for each instrument and filtered the event lists for obvious flares in the data.
We then extracted source spectra from a circular region of radius $\sim$25$^{\prime\prime}$ centered at the \chandra\ source position for each instrument. We did the same for annular background regions. The inner radius of the annulus was $\sim30^{\prime\prime}$ while the outer radius was $\sim50^{\prime\prime}$. This was to ensure that the full background region was on the same detector chip as the source region and was not contaminated by any source photons.
After generating the appropriate response files using {\tt rmfgen} and {\tt arfgen}, we merged all four spectra (MOS 1 and 2 for each observation) using the SAS command {\tt epicspeccombine}. Finally, we binned the source spectra, again, by an SNR$>2$. We fit the \xmm\ spectrum in the 5 -- 10 keV range. Above 10 keV, the EPIC cameras' effective areas sharply decrease (Refer to the \xmm\ Users Handbook Issue 2.22, 2024). Below 5 keV, there was only a single bin in the \xmm\ spectrum and it was contaminated by high background.  

\subsubsection{\nustar}
There were five \nustar\ observations whose pointing coordinates were within $8^\prime$ of NGP 21's \chandra\ coordinates. Despite the relatively small offset between NGP 21's position and the pointings of the observations, NGP 21 was only in the field of view of one of those observations (ObsID 40031002001). 

There were 29 \nustar\ observations available within $8^\prime$ of NGP 63's \chandra\ coordinates. In order to ensure the quality of the extracted spectra, however, we only analyzed the three observations with exposure times greater than 100 ks (ObsIDs 30001002001, 30502014002, and 90501330002). 

The Level 1 science event files for the \nustar\ observations were reduced using NUSTARDAS v1.9.2. Source and background regions were taken using the FPMA/B Level 2 event files. The source regions were circular and centered, again, on the \chandra\ coordinates. The background regions were circular offset from the source but on the same detector chip. Using these regions, we ran {\tt nuproducts} for FPMA/B to create source and background spectra as well as response matrices. {\tt Nuproducts} utilized the CALDB version 20220331. 

Like the \chandra\ and \xmm\ spectra, the FPMA and FPMB spectra for NGP 21 were binned by an SNR$>2$. The \nustar\ spectra for NGP 63, however, suffered from high background contamination and could not be binned by its SNR. To improve the quality of the spectrum, we merged the spectra with the highest count rates (30001002001 FPMA, 90501330002 FPMA, 90501330002 FPMB) using HEASoft's {\tt addspec}. Despite the improved count rate, the merged spectrum was still grouped by a minimum of 20 counts per bin. 

The \nustar\ data for NGP 21 were fit in the range of 3 -- 24 keV. The \nustar\ data for NGP 63, though, were fit between 8 -- 24 keV due to the high background contamination at lower energies. The 24 keV cutoff was chosen since \nustar\ background is dominated by instrumental effects above that energy \citep{Wik2014}. 
An apparent "bump" is visible in NGP 63's \nustar\ spectrum near $\sim12$ keV.  This bump is most likely an artifact of statistical fluctuations due to the high background and poor signal-to-noise ratio in the NGP 63 \nustar\ data.

\subsection{Spectral Analysis Results}

We fit each source to three phenomenological models using the XSPEC spectral fitting package: a power-law model, a power-law model with a Gaussian component, and a thermal APEC model. Each model was multiplied by {\tt tbabs}, an absorption component in order to account for ISM absorption. The {\tt tbabs} component utilized the Wilms abundance data \citep{Wilms2000}. Additionally, a constant cross-normalization factor was applied to each model with \chandra's factor frozen to 1. The unabsorbed flux was calculated using the {\tt cflux} convolution model in the 2--24 keV range. The resulting luminosity was calculated assuming a distance of 8 kpc. Table \ref{tab:spectra} reports the results of these spectral models for NGP 21 and 63.

\setlength{\tabcolsep}{20pt}
\renewcommand{\arraystretch}{1.2}
\begin{table}[]
\begin{center}
    \caption{Results from joint \chandra, \nustar\ and (for NGP 63) \xmm\ spectral fitting for NGP 21 and NGP 63.  Listed errors correspond to $1\sigma$ uncertainties.}\label{tab:spectra}
\begin{tabular}{c|cc|cc}
\hline\hline
Source
&
\multicolumn{2}{|c|}{\textbf{NGP 21}} & \multicolumn{2}{|c}{\textbf{NGP 63}}
\\
\hline
Parameter & \tt{pow} & \tt{APEC}
& \tt{pow} & \tt{APEC} \\
\hline
$N_H$ [$10^{22}$ cm$^{-2}$)]	& $49.5_{-14.8}^{+17.1}$	&	$44.9_{-10}^{+12}$	&	$18.5_{-8.8}^{+9.3}$	&	$30.8_{-4.8}^{+5.1}$	\\
$\Gamma$	&	$1.9\pm0.8$	&	...	&	$0.3\pm0.6$		&	...	\\
$kT$ (keV)	&	...	&	$14.5_{-7.8}^c$	&		...	&	$43.1_{-19.0}^c$	\\
$Z$ [$Z_\odot$]	&	...	&	$0.5_{-0.5}^c$	&		...	&	$5.0_{-2.7}^c$	\\
$C^{a}$ (\nustar/\xmm) & 1.0/1.5 & 1.0/1.5 & 0.2/1.0 & 0.4/1.0 \\
$\chi_{\nu}^{2}$ (dof)	&	0.91 (40)	&	0.89 (39)	&	1.01 (163)		&	1.01 (162) \\ 
$f_x$ [$10^{-13}$ \fluxcgs] (2-24 keV)	&	$6.8_{-1.3}^{+2.6}$	&	$3.2_{-0.8}^{+1.2}$ 	&	$2.7_{-0.4}^{+0.6}$	&	$3.6\pm0.5$ \\
$L_x\,\!^{b}$ [$10^{33}$ \lumcgs] (2-24 keV)	&	$5.2_{-1.0}^{+2.0}$	&	 $2.5_{-0.6}^{+0.9}$	&	$2.1_{-0.3}^{+0.4}$	&	$2.7_{-0.3}^{+0.4}$ \\[2pt]
\hline 
\end{tabular}
\end{center}
\vspace{-0.2cm}
$^a$Range of Constant factors with \chandra\ spectrum frozen to 1.\\
$^b$Assuming a distance of 8 kpc.\\
$^c$Upper bound of confidence interval reached hard limit in XSPEC ($kT=64$ keV and $Z=5Z_\odot$).\\
\end{table}

\begin{figure}
    \centering
    \includegraphics[width=15cm]{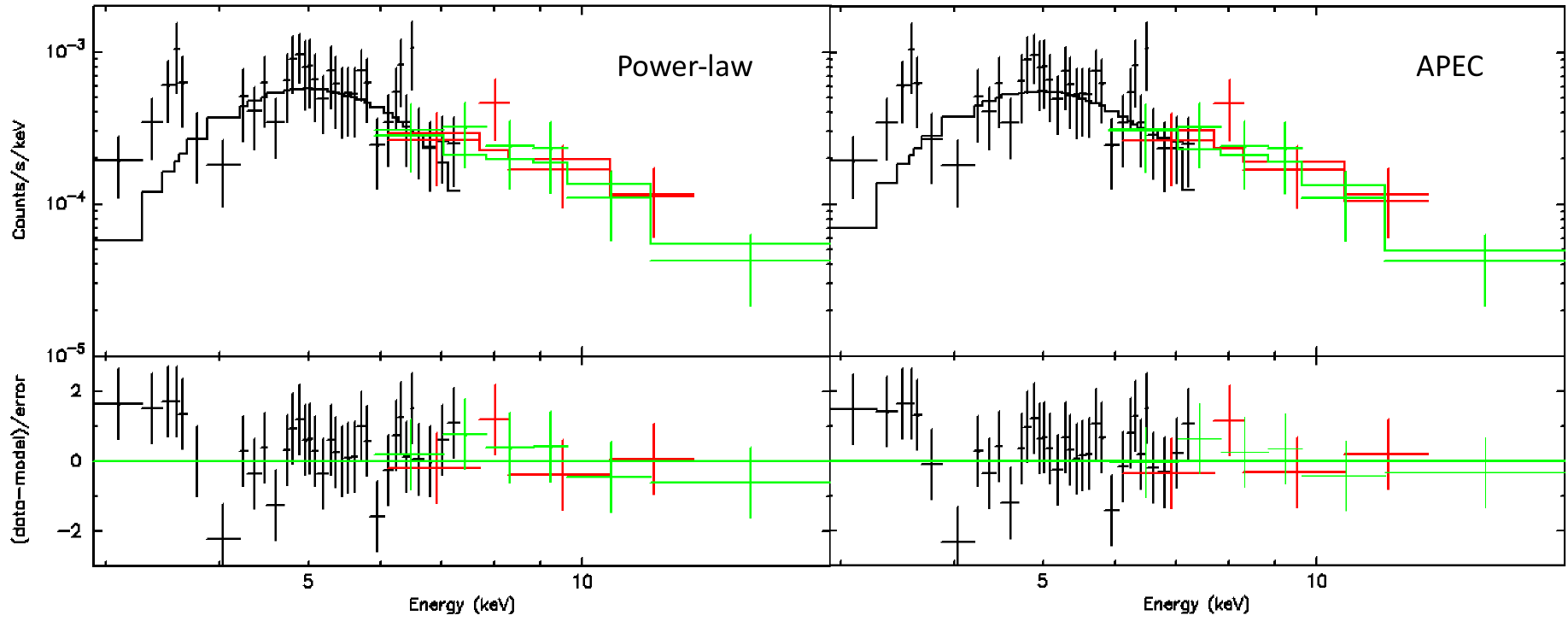}
    \caption{NGP 21 \chandra\ (black) and \nustar\ FPMA/B (red and green, respectively) spectra and residuals fit to an absorbed power-law model (left) 
    and thermal APEC model (right).  }
    \label{fig:ngp21spectra}
\end{figure}
\begin{figure}
    \centering
    \includegraphics[width=15cm]{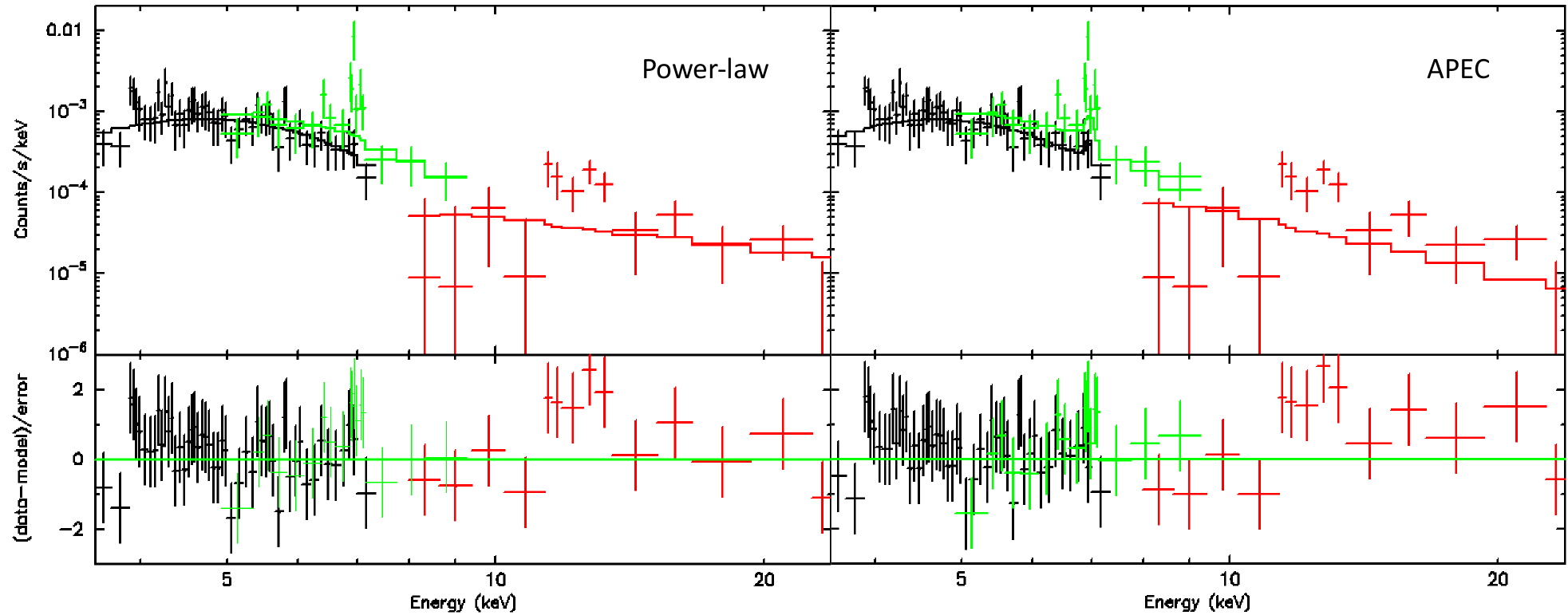}
    \caption{NGP 63 \chandra\ (black), \xmm\ (green) and \nustar\ (red) spectra and residuals fit to an absorbed power-law model (left) and thermal APEC model (right).}
    \label{fig:ngp63spectra}
\end{figure}
\subsubsection{NGP 21}
The merged \chandra\ spectrum for NGP 21 contained 190 net counts and a $\sim101$ ks exposure time. The two \nustar\ spectra added an additional 119 net counts from a combined exposure time of $\sim80$ ks.
The spectra were relatively well fit to an absorbed power-law model ({\tt constant*tbabs*powerlaw} in XSPEC) with a $\chi^2$ of 36.4 with 40 degrees of freedom (dof). The best-fit column density, $N_H=49.5_{-14.8}^{+17.1}\times10^{22}$ cm$^{-2}$, is much higher than the average column density along the line of sight, $4.36\times10^{21}$ cm$^{-2}$ \citep{HI4PI2016}. A high column density may be expected in the case of an HMXB due to the stellar wind accretion they can exhibit. This is also the case in an IP where the accretion curtains may cause greater absorption. It is also possible, however, that the relatively few low-energy bins in the NGP 21 spectra are simply leading to a poor constraint on the column density. The source is also best-fit with a photon index $\Gamma=1.9\pm0.8$. The resulting unabsorbed flux and luminosity are $6.8_{-1.3}^{+2.6}\times10^{-13}$ \fluxcgs and $5.2_{-1.0}^{+2.0}\times10^{33}$ \lumcgs, respectively.

The addition of a Gaussian line component to the model (now, {\tt constant*tbabs*(powerlaw+gauss)} in XSPEC) did not improve the fit ($\chi_\nu^2 = 0.76$). Both the column density and photon index were fit to significantly different values at $N_H=0.09_{-0.09}^{+18}\times10^{22}$ cm$^{-2}$ and $\Gamma=-0.3_{-1.7}^{+1.2}$. Again, the column density is poorly constrained. 

The Gaussian line component is similarly not well fit. Although the line energy is centered at $E_{line}=7.0\pm0.6$ keV, coinciding with the 6--7 keV Fe-line complex typical of many X-ray spectra, the line width is poorly constrained. With $\sigma_{line}=1.8_{-0.5}^{+0.8}$ keV and $EW=18.7_{-6.9}^{14.1}$ keV, the Gaussian component is likely fitting to the full continuum rather than an individual emission line. Furthermore, the low $\chi^2$ of 28 with 37 dof indicates 
that the spectrum is being over-fit with too few bins.

NGP 21 was finally fit with a thermal APEC model ({\tt constant*tbabs*apec} in XSPEC). The fit was of similar quality to the initial power-law fit with a reduced $\chi^2$ of 0.89. The column density was not significantly different from that of the power-law fit. However, the best fit temperature and abundance were both poorly constrained with the upper limit on the temperature's confidence level reaching the maximum allowed in XSPEC (64 keV) and the abundance's confidence interval spanning the full allowed range (1--5 $Z_\odot$). 

\subsubsection{NGP 63}
The merged \chandra\ spectrum for NGP 63 had 335 net counts and an exposure time of $\sim137$ ks. The merged \nustar\ spectrum contributed 242 net counts from a combined $\sim365$ ks of exposure time. Lastly, the merged \xmm\ spectrum added an additional 214 net counts over $\sim93$ ks of exposure time.
The spectra were well fit to the absorbed power-law model with $\chi^2_\nu=1.01$. The best-fit column density and photon index were $N_H=18.5_{-8.8}^{+9.3}\times10^{22}$ cm$^{-2}$ and $\Gamma=0.3\pm0.6$, 
respectively. The resulting 2--24 keV unabsorbed flux and luminosity were $2.7_{-0.4}^{+0.6}\times10^{-13}$ \fluxcgs and $2.1_{-0.3}^{+0.4}\times10^{33}$ \lumcgs. 

A clear Fe line around 6--7 keV is visible in the \xmm\ spectrum of NGP 63. This was confirmed by the well-fit power-law plus Gaussian model ($\chi^2_\nu=0.99$). Here, the line energy was best-fit to $E_{line}=6.9_{-0.04}^{+0.05}$ keV, 
consistent with H-like Fe
. While the line width was poorly constrained ($\sigma_{line}=0.02_{-0.02}^{+100}$ eV), the equivalent width was fit to $EW=270_{-120}^{+130}$ eV.  However, since this line does not appear as prominently in the \chandra\ spectrum, we cannot rule out that it is not an artifact of the diffuse Galactic background.  Such background contamination is less likely in the \chandra\ data given the much smaller source extraction aperture.

Like NGP 21, the APEC fit of NGP 63 resulted in a poorly constrained temperature and abundance. The $1\sigma$ confidence intervals for both the temperature and abundance reached the maximum values allowed in XSPEC ($kT=43.1_{-19.0}^{+20.9}$ and $Z=5.0_{-2.7}^{+0} Z_\odot$). Despite the strong fit to the thermal APEC model ($\chi^2_\nu=1.01$), failure to constrain temperature and abundance suggests that NGP 63 is a nonthermal source. \\

\section{Discussion}\label{sec:discuss}
\subsection{Source Identification}

The most common types of hard X-ray sources found in the GC region are magnetic cataclysmic variables (CVs) and low-mass X-ray binaries (LMXBs), with HMXBs being very rare.  CVs -- binary systems in which a white dwarf (WD) accretes material from a main-sequence donor -- are believed to be predominant among the population of unidentified X-ray sources \citep{Hailey2016, Anastasopoulou2023}.  Magnetic CVs can be divided into two categories: Intermediate Polars (IPs) and polars.  (Unlike their magnetic counterparts, non-magnetic CVs typically do not exhibit hard X-ray emission.)  Polars have very powerful magnetic fields ($B>10$ MG) that prevent the formation of an accretion disk, and their WD spin is synchronized with the orbital period \citep{Mukai2017}. As implied by the name, IPs have slightly weaker magnetic fields than Polars ($B\sim0.1-10$ MG) \citep{Ramsay2008}.  The magnetic field disrupts the inner part of the accretion disk; the accretion flow travels from the truncated inner edge along the magnetic field lines, and is eventually funnelled into an accretion column above the WD magnetic pole, where it hits a standing shock and rapidly decelerates before reaching the WD surface \citep{Hailey2016, Mukai2017}.  
X-ray spectra of CVs often show an Fe complex with emission lines at 6.4 keV (fluorescent), 6.7 keV (He-like), and 6.97 keV (H-like) \citep{Mukai2017, Vermette2023}.  Typical IP plasma temperatures range between $kT\sim20-40$ keV \citep{Vermette2023}, while polars are usually around $kT\sim5-20$ keV \citep{Mukai2017}.  
The Fe line equivalent width (EW) can sometimes be used to distinguish between different types of CVs. The mean 6.7 keV Fe line EW for IPs, polars, and dwarf novae (DNe) are $107 \pm 16$, $221 \pm 135$, and $438 \pm 84.6$ eV, respectively \citep{Xu2016}. 

In HMXBs and LMXBs, the accreting compact object is a neutron star (NS) or black hole (BH) rather than a WD.  Like CVs, HMXBs and LMXBs can exhibit Fe emission lines, but these are typically narrower ($\sigma_{\rm line}<150$ eV) and centered at $\sim 6.4$ keV, corresponding to the Fe fluorescence line \citep{Gimenez-Garcia2015}.  HMXB continuum spectra are usually hard and non-thermal ($\Gamma \sim1-2$).  NS-LMXBs in quiescence often -- though not always -- show predominantly soft thermal X-ray emission, while quiescent BH-LMXBs tend to have non-thermal spectra ($\Gamma \sim1.5-2.5$; \citet{Mori2021}).

In evaluating the X-ray spectra of our HMXB candidates, we pay special attention to the properties that can allow us to distinguish between CVs, LMXBs, and HMXBs -- namely, their spectral shape, hardness, and Fe emission line complex.  Given the low probability of a random overlap with the bright NIR counterpart candidates we identified ($>3\sigma$ for a "true" association), it is likely that these systems do indeed have high-mass donors.  This is especially the case for NGP 63, 
which was previously identified as a red supergiant (RSG) X-ray binary \citep{Gottlieb2020} in a follow-up study of \citet{Dewitt2013}, who had flagged its \chandra\ counterpart (XID 6592) as a candidate symbiotic binary.   
NGP 63 boasts 
an expected number of random matches $\EX_{\rm rand}\simlt10^{-4}$, 
strongly indicating that the overlapping RSG is its "true" counterpart.  
For NGP 21, we cannot fully rule out that the NIR counterpart candidate is an evolved (red giant) low-mass star; in particular, both its color and magnitude are consistent with a Red Clump star.  However, compact objects with evolved donors (e.g. symbiotic binaries) accrete at much higher rates than those with main sequence donors, which commonly results in frequent bright outbursts \citep{Lin2019}.  No such outbursts have been observed for NGP 21; NGP 63 has in the past shown bright ($L_{\rm X}\sim10^{38}$ \lumcgs) X-ray flares and significant variability in the NIR, consistent with a supergiant fast X-ray transient (SFXT\footnote{\citet{Gottlieb2020} notes that no other SFXT has previously been identified with an RSG donor.}) \citep{Gottlieb2020}.  We therefore consider it unlikely that either source is an LMXB -- assuming that the NIR sources we identified are the true counterparts to these X-ray sources.  

Although a random association with such bright NIR sources for NGP 21 and especially NGP 63 is unlikely, it cannot be ruled out completely.  Given the estimated density of faint, low-mass stars with $K\simgt20$ at the GC, virtually every X-ray source in the region is bound to overlap with a low-mass star along our line of sight.  Our determination that these X-ray sources are HMXB candidates rests not on the lack of detection of such overlapping faint NIR sources (due to the extreme crowding that prevents their detection), but on the low probability that the bright NIR sources that have been observed overlapping the X-ray position are randomly aligned.  We therefore conclude that these NIR sources are most likely the true counterparts.

The biggest remaining question, then, is whether they are high-mass binaries with a WD ($\gamma$ Cas-type; \citet{Motch2015}) or BH/NS (HMXB) primary.  We find that the X-ray spectra of both NGP 21 and NGP 63 are more consistent with the HMXB scenario, given their hard, non-thermal spectra; temperatures could not be constrained when a thermal fit was attempted.    
Nevertheless, we cannot definitively rule out a Be-WD $\gamma$ Cas-type binary.  
Time-resolved NIR spectroscopic follow-up observations may enable us to confirm the association with the X-ray sources, and possibly even determine the nature of the compact object.  
This is true also for the other five HMXB candidates, which have higher random association probabilities.  We cannot rule out that those sources, too, are HMXBs.  We leave it to future in-depth studies to resolve the question of their identification.

\subsection{Implications of Detecting Faint HMXBs in the GC}

Understanding X-ray binaries' luminosity function can provide insight into their evolution. Previous studies of the HMXB luminosity function tended to focus on the higher-luminosity end of that function ($L_{\rm X}\simgt10^{35}$ \lumcgs; \citet{Grimm2003, Zuo2014, Mineo2012, Misra2023}).  
Galactic HMXBs have been analyzed at slightly fainter levels, with their luminosity function studied down to $\sim10^{34}$ \lumcgs \citep{Lutovinov2013}. While these higher--luminosity studies are significant for their insight into topics such as star formation rates and binary evolution, there is a clear gap in understanding the lower--luminosity end of this function. 
The increased sensitivity of high-energy X-ray telescopes like \nustar\ offers a unique opportunity to probe this open question. 
Given the slope of the luminosity function, the discovery of low luminosity HMXBs may yield 
more BH-HMXB systems in general, and Be-BH systems in particular, both of which are very rare and tend to have fainter quiescent luminosities than their NS counterparts \citep{Munar-Adrover_2014, Fornasini2016PhDT}. 
Be-BH systems are possible progenitors of NS-BH systems, which themselves are sources of gravitational waves. Therefore, understanding the fainter HMXB population may allow us to better constrain rates of gravitational wave events.

Apart from its general importance towards understanding the luminosity function of HMXBs, the discovery of faint HMXBs in the GC 
is of particular significance.  While the CMZ hosts the largest concentration of LMXBs observed anywhere in our Galaxy \citep{Mori2021}, very few HMXBs have hitherto been identified in the central degrees, and only one (NGP 63; \citet{Gottlieb2020}) within 100 pc of Sgr A* -- a region containing dozens of LMXBs, $>10\%$ of all LMXBs observed in the Galaxy \citep{Liu2006, Bird2016, Mori2021, Fortin2023, Fortin2024}.  The dearth of HMXBs in the CMZ is particularly surprising given the abundance of young massive stars in the region, an optimal environment in which to find the necessarily short-lived HMXBs \citep{Hatchfield2024}.  A hidden population of fainter HMXBs could explain the apparent deficiency in identified HMXBs in the CMZ region.

\subsection{Probabilities of Missed 
Sources: Upper Limits on the Number of HMXBs in the GC} \label{subsec:mmprob}

As described in Section \ref{subsec: rprob1}, the probability of identifying a NIR counterpart to an X-ray source depends primarily on the magnitude of the NIR source and the angular separation $\phi$ between the X-ray and NIR source positions.  For our analysis, we limited our search area for NIR counterparts to the 95\% \chandra\ uncertainty region associated with each X-ray source.  

Whereas in Section \ref{subsec: rprob1} we concerned ourselves with the probability that a \emph{detected} NIR source is a real counterpart to an X-ray source, vs randomly overlapping ("false-positive counterparts"), in this section, we examine the probability that a real NIR HMXB counterpart was \emph{not} detected ("false negatives").  This could have occurred either due to incompleteness of the NIR data, or because the NIR counterpart was located outside of the 95\% \chandra\ X-ray uncertainty region we considered.  By determining the number of probable HMXBs that were not identified in our survey, we can place constraints on the total HMXB population in the GC overall.

We begin by evaluating the fraction of X-ray/NIR 
matches that are "true" counterparts vs randomly overlapping.  We conducted a series of simulations that allowed us to constrain the expected number of random matches as a function of $K$, as described below.

First, we 
collated both the \chandra\ X-ray sources (from CSC2.1) and the  VVV NIR sources within the \nustar\ survey 
region ($-0.9<l<0.8$; $-0.15<b<0.35$ deg).  We cross-checked each X-ray source against all of the NIR sources to find the total number of observed astrometrically aligned\footnote{We consider two sources astrometrically aligned if their respective position uncertainty regions overlap.} sources, $N_{\rm tot}$.  We then shifted each of the X-ray sources by a random offset to its original position, and repeated the cross-matching process to see how many NIR sources would randomly overlap with the simulated X-ray source positions.  
We set the random offset maximum at $\sim15$\asec to avoid disturbing the overall source density distribution.  
To ensure a statistically robust expected number of random matches estimate, we repeated the random-shift-and-cross-match process 1000 times, obtaining a distribution of random match probabilities of which we extracted the mean, $\overline{N}_{\rm rand}$, and median, $\tilde{N}_{\rm rand}$.  

Since the number of randomly overlapping NIR sources is determined by the overall density of such sources, and density varies by luminosity, we repeated the random match simulations outlined above for NIR sources with a range of magnitudes thresholds, obtaining a range of values $\overline{N}_{\rm rand}(K)$ and $\tilde{N}_{\rm rand}(K)$.  We found that the latter was a better representation of the expectation value for the distributions $N_{\rm rand}(K)$ at brighter magnitudes where both real and random matches were rare and $N_{\rm rand}$ peaked at 0.    
We then subtracted $\tilde{N}_{\rm rand}(K)$ and, separately,
$\overline{N}_{\rm rand}(K)$ -- which we take as the expectation value for the number of random matches at a given magnitude $K_i$ -- from the number of NIR sources \textit{at the same magnitude} that were observed to be astrometrically aligned, $N_{\rm tot}(K)$.  The difference of $N_{\rm tot}(K)-\tilde{N}_{\rm rand}(K)$ (or alternatively, $N_{\rm tot}(K)-
\overline{N}_{\rm rand}(K)$) was taken to be the number of "real" counterparts, $N_{\rm real}(K)$.  

We then obtained the fraction of "real" X-ray/NIR matches by taking $f_{\rm real}(K)=N_{\rm real}(K)/N_{\rm tot}(K)$.
Conversely, we can calculate the fraction of observed matches that are likely random by taking $f_{\rm rand}(K)=\overline{N}_{\rm rand}(K)/N_{\rm tot}(K)$.   

We then obtained the \emph{fraction of X-ray sources with a "real" NIR counterpart} by taking $F_{\rm real}(K)=N_{\rm real}(K)/N_{\rm X}$, where $N_{\rm X}$ is the total number of X-ray sources within the survey region.  
($F_{\rm real}(K)$ -- describing the fraction of \emph{X-ray sources} with real NIR counterparts -- should not be confused with $f_{\rm real}(K)$ as defined above, which represents the fraction of observed \emph{matches} that are real).  
This empirically determined fraction $F_{\rm real}(K)$ can be used to evaluate the probability of an X-ray/NIR pair \emph{not} being identified due to the NIR source being located outside the above-defined X-ray uncertainty region.

We can define the probability of a "real" NIR counterpart not being identified as 
\begin{equation} \label{eq:phix}
    P(\phi, K) = F_{\rm real}(K)\times \int_{2\sigma}^{\infty}\!\!p(\phi)d\phi
\end{equation}
where $F_{\rm real}(K)$ is the fraction of X-ray sources with "real" NIR counterparts of magnitude $K$ (as described above) and $\int_{2\sigma}^{\infty}p(\phi)d\phi$ (where $p(\phi)$ is defined by equation \ref{eq:phi}) is the probability of a "real" NIR counterpart being located outside a radius $2\sigma$ from the X-ray position centroid.  For the negative match probability estimate using equation \ref{eq:phix} above, we adopted $\sigma=\sigma_{\rm X}$, since the NIR position uncertainties tend to be much smaller ($\sigma_{\rm IR}\simlt0.1\asec$).  For the purposes of this calculation, we assumed that the X-ray uncertainty region is circular and of radius $2\sigma_{\rm X}=0.5\asec$, which is typical for 95\% \chandra\ uncertainties.  The probability density function $p(\phi)$ was normalized such that $\int_{0}^{\infty}p(\phi)d\phi=1$.  We integrated equation \ref{eq:phix} from $2\sigma$ to infinity; the lower limit was selected to reflect the fact that our NIR match search was conducted using the 95\% \chandra\ position uncertainties.  $P(\phi, K)$ for a range of magnitudes $K$ 
can be found in Table \ref{tab:f_k}.

\setlength{\tabcolsep}{12pt}
\begin{table}
\caption{Random match probability (RMP): total number of observed matches $N_{\rm tot}$, mean number of random matches $\overline{N}_{\rm rand}$ over 10000 simulations, and the projected number of "real" matches, $\overline{N}_{\rm real}=N_{\rm tot}-\overline{N}_{\rm rand}$ at magnitudes $K<K_{\rm lim}$.  We list the fraction of X-ray sources that have genuine 
NIR counterparts at magnitude brighter than $K_{\rm lim}$: $F_{\rm real}(K)=\overline{N}_{\rm real}/N_{\rm X}$, where $N_{\rm X}$ is the total number of X-ray sources.    
Finally, $f_{\rm rand}(K)$ denotes the fraction of matches with $K<K_{\rm lim}$ that appear to be random, while $P(\phi, K)$ is an estimate of the probability that a true NIR counterpart failed to be identified due to being located beyond the $\sim2\sigma$ limit.
\label{tab:f_k}}  
\begin{tabular}{l c c c c c c c} 
\hline\hline
$K_{\rm lim}$ & $N_{\rm tot}$ & $\overline{N}_{\rm rand}$ & $\overline{N}_{\rm real}$ & $f_{\rm real}(K)$ & $F_{\rm real}(K)$ & $f_{\rm rand}(K)$ & $P(\phi, K)^*$ \\  [0.11ex]
[mag] &   &   &   &  &  &  \\ [1ex] 
\hline
10.8   &   1   &   0.09   &   0.91   &   0.91   &   0.05   &   0.09   &   0.01  \\
11.2   &   2   &   1.12   &   0.88   &   0.44   &   0.05   &   0.56   &   0.01  \\
11.6   &   4   &   1.69   &   2.31   &   0.58   &   0.12   &   0.42   &   0.02  \\
12.0   &   5   &   2.16   &   2.84   &   0.57   &   0.15   &   0.43   &   0.03  \\
12.4   &   6   &   2.78   &   3.22   &   0.54   &   0.17   &   0.46   &   0.03  \\
12.8   &   8   &   3.62   &   4.38   &   0.55   &   0.23   &   0.45   &   0.05  \\
13.2   &   9   &   4.6   &   4.4   &   0.49   &   0.23   &   0.51   &   0.05  \\
13.6   &   11   &   5.41   &   5.59   &   0.51   &   0.29   &   0.49   &   0.06  \\
14.0   &   12   &   6.57   &   5.43   &   0.45   &   0.29   &   0.55   &   0.06  \\
14.4   &   12   &   7.52   &   4.48   &   0.37   &   0.24   &   0.63   &   0.05  \\
14.8   &   13   &   8.28   &   4.72   &   0.36   &   0.25   &   0.64   &   0.05  \\
15.2   &   15   &   9.14   &   5.86   &   0.39   &   0.31   &   0.61   &   0.06  \\
15.6   &   15   &   9.51   &   5.49   &   0.37   &   0.29   &   0.63   &   0.06  \\
16.0   &   15   &   9.67   &   5.33   &   0.36   &   0.28   &   0.64   &   0.06  \\
\hline 
\end{tabular} \\ [3 pt]
$^*$For the purposes of these calculations, we assumed that $\phi=0.5$.  
\end{table}

At a magnitude of $\sim16.5$, which corresponds to the low-luminosity end of an HMXB donor star at the GC distance (cf. Section \ref{sec:ccorr}), we find that $F_{\rm real}(K)\simeq0.3$.  Then, integrating Equation \ref{eq:phix} (assuming $2\sigma_{\rm X}=0.5\asec$), we get the probability of $\sim6\%$ for false negatives.  Finally, to estimate the total number of missed matches ("false negatives") $N_{\rm FN}$, we multiply Equation \ref{eq:phix} by the number of X-ray sources, $N_{\rm X}$; $N_{\rm FN}=N_{\rm X}\times P(\phi, K)$.    This indicates that 
at most there are $\sim4$ additional unidentified HMXB candidates among the remaining 70 \nustar\ sources.  Assuming that no HMXBs fall outside the \nustar\ detection limit, this would indicate a persistent lack of HMXBs in the GC; however, given the limitations of current surveys, it is entirely possible that future more sensitive observations will uncover an additional, fainter population of 
HMXBs in the CMZ. \\

\section{Conclusions}

We conducted a study of unidentified hard X-ray point sources detected in the GC with \nustar.  We searched for bright NIR counterparts consistent with HMXB donor stars.  We find that two of the sources are strong candidate HMXBs, given their probable association with bright NIR sources.  We fit the \chandra, \xmm\ and \nustar\ spectra and measure best-fit photon index $\Gamma$ and flux.  Follow-up NIR spectroscopy will be able to confirm whether the bright NIR candidate counterparts we identified are indeed HMXB donor stars, and determine their type.  The discovery of HMXBs within the CMZ region of the GC is significant for our understanding of stellar evolution in general and compact object binaries in particular.

\section{acknowledgement}
The authors gratefully acknowledge the assistance of the anonymous referee, whose thoughtful critique of an earlier draft helped us improve this paper.  
Support for this work was partially provided by the Chandra Cycle 23 Archive Data Analysis Program through SAO grant AR2-23005X. We acknowledge Orion Van Oss for assisting with X-ray spectral analysis. 
This material is based upon work supported by the National Science Foundation Graduate Research Fellowship under Grant No. DGE 2036197. GP acknowledges financial support from the European Research Council (ERC) under the European Union’s Horizon 2020 research and innovation program HotMilk (grant agreement No 865637), support from Bando per il Finanziamento della Ricerca Fondamentale 2022 dell’Istituto Nazionale di Astrofisica (INAF): GO Large program and from the Framework per l’Attrazione e il Rafforzamento delle Eccellenze (FARE) per la ricerca in Italia (R20L5S39T9). 

\bibliography{main}
\bibliographystyle{aasjournal}

\appendix

\section{Random Match Probability for NIR Counterpart Candidates}\label{sec:AB}

\begin{figure}
    \centering
    \includegraphics[width=0.5\textwidth]{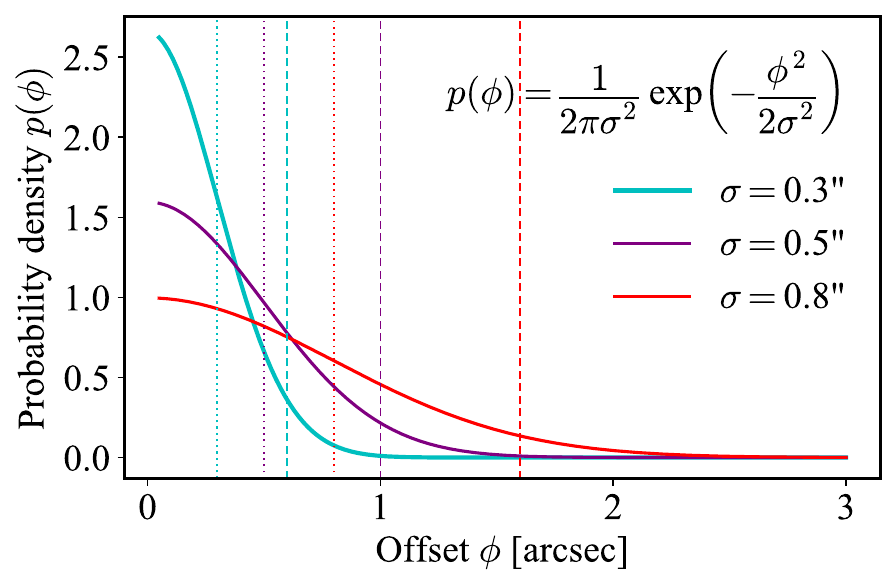}
    \includegraphics[width=0.49\textwidth]{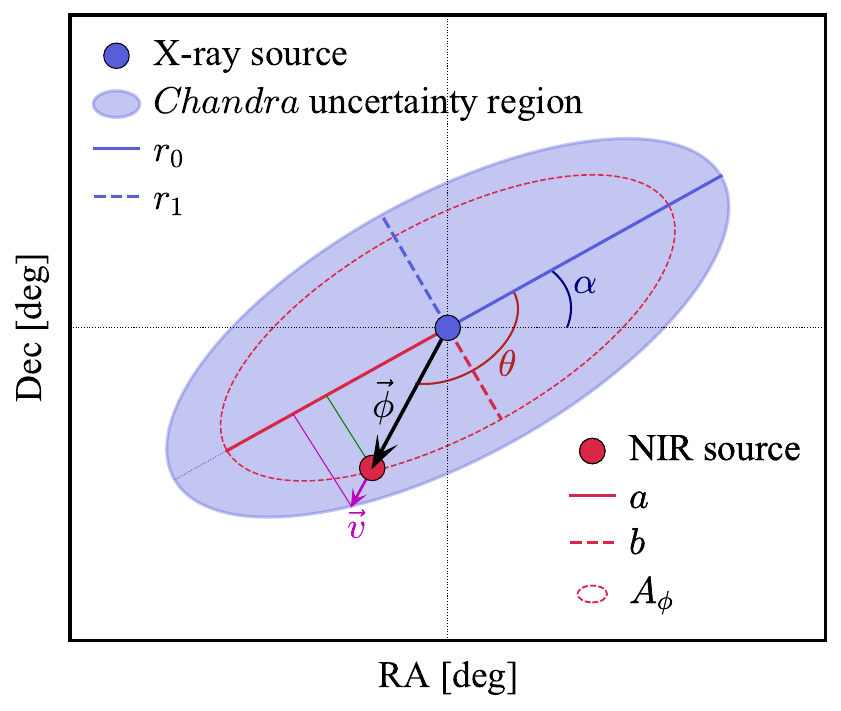}
    \caption{Left: Probability density (PDF) as a function of offset $\phi$ between the X-ray position centroid and the NIR counterpart candidate.  The PDF above was normalized such that $\int_{0}^{\infty}p(\phi)d\phi=1$.  Dotted lines mark the $1\sigma$ point containing the peak of the distribution; dashed lines denote the $2\sigma$ limit, beyond which the probability distribution drops off.  See Section \ref{subsec:mmprob} for more.  Right: diagram outlining the process for defining the region $A_{\phi}$.  See text for definitions.}\label{fig:pdf}
\end{figure}

The probability of a NIR source being associated with an X-ray source is dependent in part on the angular offset $\phi$ between the X-ray and NIR positions, as well as the uncertainties associated with those positions: $p(\phi)=\frac{1}{2 \pi \sigma^2}exp(\frac{-\phi^2}{2 \sigma^2})$.  As Figure \ref{fig:pdf} demonstrates, the probability of a true association between two sources falls off rapidly at angular distance $\phi>\sigma$, where $\sigma$ is the uncertainty of the two positions, combined in quadrature.

In order to estimate the probability of a NIR source randomly appearing at an angular distance $\phi$ from an X-ray source, we define an area $A_{\phi}$ that denotes the region containing this NIR source; we then multiply $A_{\phi}$ by the number density of NIR sources $\Sigma$ -- and divide it by the \textit{completeness} $\epsilon_i$ -- to generate an estimate for the expected number of random matches
, as given by Equation \ref{eq:rmp}.  This area $A_{\phi}$ is modelled after the \chandra\ 95\% uncertainty ellipse, which is itself defined by a semi-major axis $r_0$, a semi-minor axis $r_1$, and the angle $\alpha$ 
between the major axis of the ellipse and "local north" (which at the GC translates to the Declination axis), 
measured from North through East.   
To define $A_{\phi}$, we use  
\begin{itemize}[topsep=3pt,itemsep=0pt,partopsep=0pt, parsep=1pt]
    \item the angular offset vector $\vec\phi$ between the X-ray and NIR source positions, 
    \item the angle $\alpha$ between the semi-major axis $r_0$ of the \chandra\ uncertainty ellipse and the Declination axis, 
    \item the semi-major axis $r_0$ and semi-minor axis $r_1$ of the \chandra\ uncertainty ellipse, and
    \item the angle $\theta$ between $r_0$ and the NIR source position.
\end{itemize}
To define the semi-major axis $a$ of the elliptical area $A_{\phi}$, we first take the dot product of the vector $\vec\phi$ with the semi-major axis $r_0$, i.e. the projection of $\vec\phi$ that is aligned with $r_0$, 
$\boldsymbol{\phi_{\parallel r_0}}$.  We also calculate the projection $\bf{v_{\parallel r_0}}$ 
of a vector $\rm \vec {v}$ \textit{defined at the same angle $\theta$ with respect to $r_0$} that extends to the edge of the \chandra\ uncertainty ellipse.  The ratio of the projections  
$c=\boldsymbol{\phi_{\parallel r_0}}/\bf v_{\parallel r_0}$ provides the scaling factor  
used to calculate the semi-major axis $a$ and semi-minor axis $b$ for $A_{\phi}$: $a=c\times r_0, b=c\times r_1$.  
Finally, we adopt the inclination $\alpha$ from the \chandra\ uncertainty ellipse for $A_{\phi}$.  
Figure \ref{fig:pdf} (right) illustrates $r_0$, $r_1$, and $\alpha$ for the \chandra\ uncertainty ellipse around NGP 10, as well as $\vec\phi$, $a$, $b$, and finally, $A_{\phi}$.  The projections $\boldsymbol{\phi_{\parallel r_0}}$ and $\bf{v_{\parallel r_0}}$ are outlined in green and magenta, respectively. \\

\section{NIR Catalog Completeness}\label{sec:AC}

\begin{figure}
    \centering
    \includegraphics[width=\textwidth]{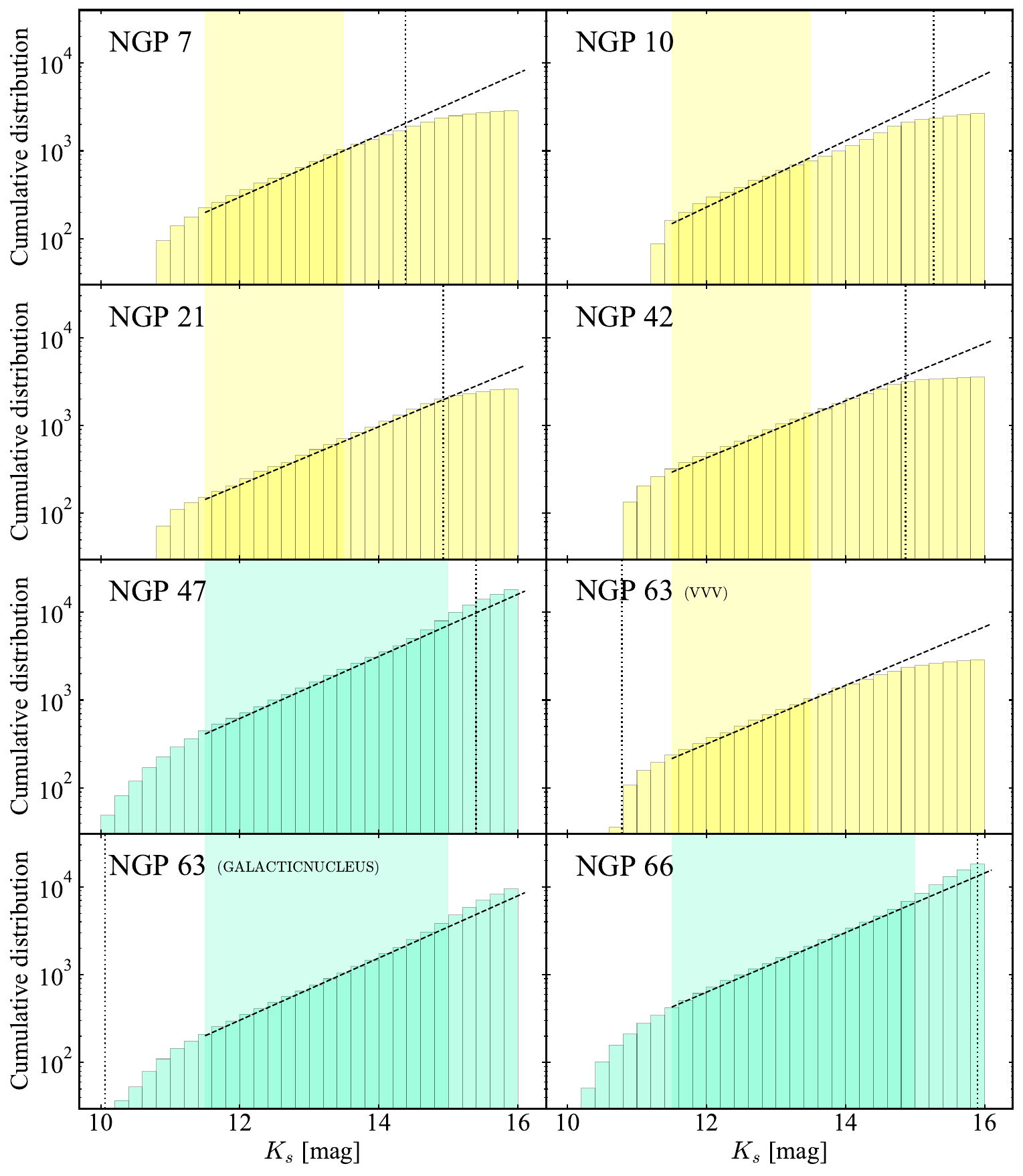}
    \caption{Distributions of NIR magnitudes in the 2\arcmin\ region around each HMXB candidate.  Vertical lines denote the magnitudes of the corresponding NIR counterpart candidate.  VVV data are shown in yellow, GALACTICNUCLEUS data in green.  The shaded region highlights the range of magnitudes in which the surveys appear complete, which we used to fit the dashed line; projected source numbers at fainter magnitudes were estimated by extrapolating the linear fit.}\label{fig:comp}
\end{figure}

If the expected number of random matches estimates $\EX_{\rm rand}$ 
were defined 
only by $\Sigma(K_i)\times A_{\phi}$ (Section \ref{subsec: rprob1}), it would rely 
on the assumption that the NIR catalogs we cross-matched our X-ray data with are complete.  This assumption would not be entirely accurate, however.  NIR source detection is hampered at the GC owing to a combination of the high stellar density, as well as the distance and foreground extinction.  Both the VVV and GALACTICNUCLEUS surveys are affected by source confusion, although to different extents and at different magnitude ranges (due to differences in angular resolution), with additional local variations within each survey (due to varying observing conditions and source density).  GALACTICNUCLEUS is more  
complete at very bright magnitudes ($K_s<11$ mag, where VVV is saturated) and at magnitudes $\simgt14$, where VVV's completeness wanes (Figure \ref{fig:comp}).   Figure \ref{fig:comp} highlights the process we used to evaluate the local completeness fraction $\epsilon$ within a 2\arcmin\ region around each HMXB candidate.   For the VVV survey, we considered $K_s = 11.5-13.5$ mag optimal for fitting the cumulative magnitude distribution, as the data appears complete within that range, with a relatively straight slope.  The GALACTICNUCLEUS survey is more complete at fainter magnitudes; therefore, we used the range $K_s = 11.5-15.0$ mag to fit the cumulative magnitude distribution.  Note that our completeness estimates may be slightly overestimated at $K_s > 15$ mag, particularly for the VVV data, 
because our straight slope fit does not account for non-exponential features in the magnitude distribution, like the Red Clump bump.  However, we do not expect that this simplification would significantly change our results, given that the more complete   
GALACTICNUCLEUS data shows evidence of only a slight excess between $15 < K_s < 16$ mag (Figure \ref{fig:comp}); additionally, most of the candidate counterparts we identified are at $K_s \simlt 15$ mag, with the exception of NGP 66, for which we have the deeper 
GALACTICNUCLEUS data.  A well-calibrated, detailed $K_s-$band luminosity function (KLF) for the GC may yield more precise values of $\epsilon$, and by extension, more accurate random match expectation values $\EX_{\rm rand}$.

\end{document}